\documentclass[twocolumn,showpacs,preprintnumbers,amsmath,amssymb]{revtex4-1}
\usepackage{textcomp}

\usepackage{graphicx}
\begin{document}

\title{Phase-separation transitions in asymmetric lipid bilayers}
\author{Shunsuke F. Shimobayashi}
\thanks{Email address}
\email{shimobayashi@chem.scphys.kyoto-u.ac.jp}
\author{Masatoshi Ichikawa}
\affiliation{Department of Physics, Graduate School of Science, Kyoto University, Kyoto 606-8502, Japan}
\author{Takashi Taniguchi}
\thanks{Email address}
\email{taniguch@cheme.kyoto-u.ac.jp}
\affiliation{Department of Chemical Engineering, Kyoto University, Kyoto 606-8502, Japan}
\date{\today}

\begin{abstract}
Morphological transitions of phase separation associated with the asymmetry of lipid composition were investigated using micrometer-sized vesicles of lipid bilayers made from a lipid mixture. The complete macro-phase-separated morphology undergoes a transition to a micro-phase-separation-like morphology via a lorate morphology as a metastable state. The transition leads to the emergence of monodisperse nanosized domains through repeated domain scission events. Moreover, we have numerically confirmed the transitions using the time-dependent Ginzburg-Landau model describing phase separation and the bending elastic membrane, which is quantitatively consistent with experimental results by fixing one free parameter. Our findings suggest that the local spontaneous curvature due to the asymmetric composition plays an essential role in the thermodynamic stabilization of micro-phase separation in lipid bilayers.  
\end{abstract}
\maketitle
%\begin{document}
%\section{Introduction}

%谷口さんの論文では、組成と曲率のカップリングによって、
%相分離したある片方がまがったほうが自由エネルギーが下がるために曲がる。
%その得と界面エネルギーの損が競合してコースニングが遅くなる。
%最終的に完全な相分離になるのかは不明である。
%柳澤さんの論文では、硬さが組成に依存するとうことで、組成と曲率のカップリングがあるといえる。
%その時にドメインが曲がることでコースニングが遅くなっている。その理由はkinteicな曲げ弾性に依ると考えられている。
%一見すると組成と曲率のカップリングでコースニングが遅くなっていると捉えれそうではある。
%が、これはkineticなものなので、谷口さんのcoarsenign delayとは分けて考えるべきである。
%kumarらの論文は、弾性係数一定で、critical off quenchでline tension とexcess areaを変えてコースニングの指数が変わるという仕事である

Generally, binary liquids relax to complete phase separation below a transition temperature.
However, in a few systems, long-range repulsive interaction between micro domains can lead to thermodynamic stable micro-phase separation~\cite{dogic,leib}.  
In terms of new material designs using membranes and the nanosized heterogeneities in cell plasma membranes~\cite{raft1,raft2,raft3}, one of the emerging problems is to elucidate 
how micro-phase separation in multi-component lipid bilayers can be stabilized and regulated.
One of the co-authors numerically predicts that the local coupling between the lipid composition and the membrane curvature can essentially lead to stable micro-phase separation~\cite{raft8}. Yanagisawa {\it et al,} have experimentally confirmed that the domain coarsening is transiently trapped due to the effective repulsive interaction between micro domains~\cite{raft7,raft9}. However, to our knowledge, the mechanism of thermodynamic stable micro phase separation is not yet fully understood~\cite{raft8,raft7,vari}. 
In this Letter, we demonstrate phase separation of a reconstituted two-component vesicle with an asymmetric lipid composition between inner and outer layers, 
ubiquitous in cell membranes.    
Moreover, we numerically investigate the transitions using a dissipative dynamical model. 

Giant unilamellar vesicles (GUVs) were prepared either electroformation or natural swelling from a mixture of 35\% DOPC, 35\% DPPC and 30\% Cholesterol at 55$^\circ$C with pure water~\cite{detail}. Vesicles with excess area can result in trapped coarsening after temperature quench~\cite{raft7}. Thus, a small amount of NaI was added to the lipid film to generate vesicles with high sphericity~\cite{hishidasan}. The average radius of the vesicles was 4.3 $\pm$ 2.0 $\mu$m (FIG. S1). Subsequently, we decreased the temperature under the miscibility transition temperature, leading to the spontaneous formation of  
liquid-ordered L$_\text{o}$ domains (DPPC rich) in a liquid-disordered L$_\text{d}$ matrix (DOPC rich), further coarsening to the fully phase-separated state through a collision-coalescence process. Simultaneously, GM1 was dissolved in pure water and then ultrasonicated for two hours~\cite{detail}. A dynamic light scattering measurement indicates the formation of micelle structure ($\sim$ 6 nm) made of GM1 molecules ($\sim$ 2 nm) in several tens of minutes after the ultrasonication (FIG. S2). Subsequently, 0.25 mM DOPC/DPPC/Chol solution was mixed with an equal amount of 0.25 mM GM1 solution. Time development of the average intensity of GM1 was measured along the equatorial orbit of each GUV using confocal microscopy, indicating that a GM1 insertion process into the GUV is equilibrated in approximately ten minutes (FIG. S3)~\cite{GM1,height1,height2}.  
Then, GM1 preferentially partitions into L$_\text{o}$ phase (FIG. S4). According to the intensity measurement of seven samples by the microscopy, the ratio of GM1 concentration in L$_\text{d}$ phase to that in L$_\text{o}$ phase is estimated to be 0.49 $\pm$ 0.09. The average mole fractions of GM1 in GUVs are estimated to be $2.90\pm0.74$\% in L$_\text{d}$ phase and $6.14\pm1.70$\% in L$_\text{o}$ phase, respectively, from comparisons between the mean intensities of GM1 prepared in the film and inserted from the outside (FIG. S5). The following phase-separation transition was identified by the difference of fluorescent intensities amongst the dyes used. The L$_\text{d}$ phase, L$_\text{o}$ phase, and GM1 were stained by rhodamine-DOPE (0.5\%), NBD-DPPE (1\%), and Bodipy-GM1 (0.6-1\%), respectively~\cite{detail}.    

%\begin{figure}[htbp]
%\centering
%\includegraphics[width=75mm]{figure1ver2.eps}
%\caption{\label{1}{ GM1 distribution on phase-separated surface.} GM1 preferentially partitions into L$_\text{o}$ phase. The GM1 ratio in Ld phase to that in Lo phase is estimated to be $0.49\pm0.09$ by using seven samples. The scale bar is 5 $\mu$m.   
%}
%\end{figure}
\begin{figure}[htbp]
\centering
\includegraphics[width=75mm]{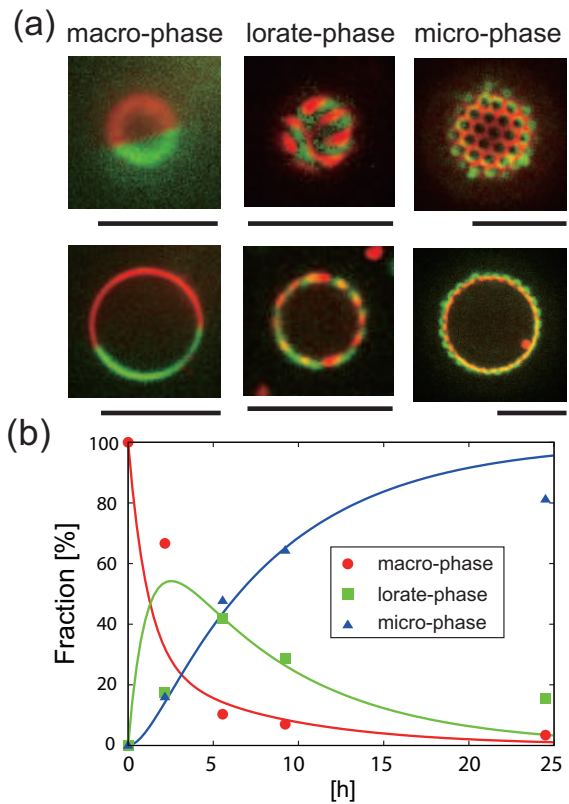}
\caption{\label{2}{ Phase-separation transitions.} (a) Confocal microscope images of top (upper) and cross-sectional (lower) views of GUVs with three different morphologies (macro-phase, lorate-phase and micro-phase). Red and green represent L$_\text{d}$ and L$_\text{o}$ phases, respectively. The scale bars are 10 $\mu$m. (b) Time variation of the fractions of vesicles exhibiting each one of these three phases. The symbols denote the experimental results and the solid lines are theoretically fitted ones.} 
\end{figure}

%It is often observed that the angle of bifurcation of a L$_\text{o}$ domain into two lorate domains are showed to be approximately 120$^\circ$ in agreement with previous results, however the lorate domains do not coalesce in contrast to the same previous results~\cite{webb}.

\begin{figure}[htbp]
\centering
\includegraphics[width=75mm]{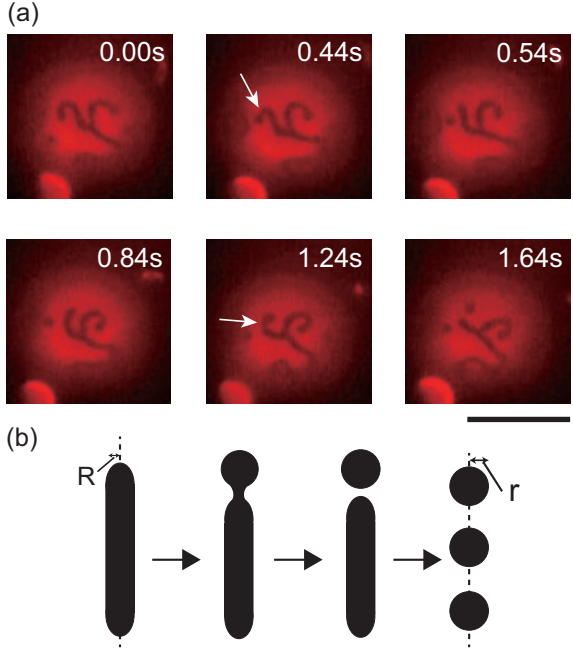}
\caption{\label{3}{ Domain scissions observed in the transition from lorate-phase to micro-phase.} (a) The white arrows indicate narrow constrictions emerging immediately before each scission event. The times are indicated for each image. The scale bar is 5 $\mu$m. (b) Schematic illustration of domain scission. $R$ and $r$ represent the length of the minor axis of lorate domains and the radius of generated circular domains, respectively.   
}
\end{figure}

Typically, three different types of phase separation morphologies are observed: (a) a fully phase separated morphology (hereafter referring as ``macro-phase"), (b) a lorate morphology (``lorate-phase") and (c) a micro-phase-separation-like morphology (``micro-phase"). L$_\text{o}$ domains in the lorate-phase and the micro-phase slightly bud toward the outside of the vesicles.
At low L$_\text{o}$ domain densities, they diffuse within a fluid L$_\text{d}$ matrix (Movie S1 and S2). On the other hand, at high densities, they organize into hexagonal order on the spherical surface, and hardly move (FIG. 1(a)). The time development of the fraction of vesicles exhibiting each one of these three phases are shown in FIG. \ref{2}(b), where GUVs displaying a coexistence of two different phases were discarded. All the vesicles initially exhibit a fully phase-separated state and subsequently the fraction of the macro phase state decreases continuously towards zero. On the other hand, the fractions of the lorate- and micro-phases increase from zero. The former reaches its maximum after about five hours and then decreases, but the latter continues to increase. These results suggest that the macro-phase undergoes a transition to the micro-phase via the lorate-phase as a metastable state. Therefore, the successive transition of state can be described by two coupled differential equations so that the experimental results are fitted as shown in FIG. 1(b) using the Levenberg-Marquardt method (Supplementary materials). The interface between L$_\text{d}$ and L$_\text{o}$ phases are rapidly transformed at the transitions. 
The interface transformation at the transition from the macro- to lorate-phase could not be captured due to the fluorescence which suppresses it, possibly due to the photo-induced oxidation of lipids~\cite{oxid}. On the other hand, we have succeeded in capturing the transformation from lorate- to micro-phase. A narrow constriction at a tip of a lorate domain emerges and immediately gives rise to domain scission (time scale less than 1 s; see white arrows in FIG. 2(a) and Movie S3). The scission event is repeated, which leads to the complete transformation from lorate domains to monodisperse micro domains. 
The ratio of the length of the minor axis of lorate domains, denoted by $R$, to the radius of micro domains, $r$ is the selected dimensionless wave number. This was roughly estimated to be approximately 0.7, which satisfy the Plateau's prediction that the selected wave number is less than 1 (FIG. \ref{3}(b)).
However, the instability should be related to not only the dissipation of the interface energy, known as Rayleigh-Plateau instability, 
but also that of the bending energy of the membrane with non-zero curvature.  
Further investigation of this instability represents intriguing future work. 
The size of the circular domains, $2r$, were measured as a function of the L$_\text{d}$ matrix radius, denoted by $R_{1}$ (FIG. 3). The solid line represents the best fit by a linear least-squares method, whose slope and intercept are 0.20 $\pm$ 0.06 and 0.20 $\pm$ 0.17 $\mu$m, respectively. The results reveal the emergence of nano-sized L$_\text{o}$ domains near the limit of optical resolution and a linearly dependent relationship between $2r$ and $R_{1}$, implying similarity of shape (FIG. 3). 
When a GM1 solution of increased concentration is added, we could not detect the micro domains, indicating that the domain size decreases below the optical limit.          
These findings strongly suggest that even a slight asymmetry of lipid distribution between inner and outer leaflets has the potential to dramatically change phase separation morphologies in lipid bilayers.

%To reduce geometrical errors of distance measurement coming from the curvature of the vesicle surface in the image plane, the domain sizes were measured near the top of vesicles.
%These events should be Rayleigh-Plateau type instability in the one-dimensional interface.
% $dx/dt=-k_{1}x+k_{2}y$ and $dy/dt=k_{1}x-k_{2}y-k_{3}y+k_{4}(1-x-y)$, where $x$, $y$ and $1-x-y$ represent the fractions of the macro-, lorate- and micro-phases, respectively. $k_{1}$, $k_{2}$, $k_{3}$, and $k_{4}$ show the transition rates from the macro-phase to the lorate-one, vice versa, from the lorate to the micro, and vice versa, respectively. The results can be fitted as shown in FIG. 3(a), by substituting the values $k_{1}=0.75 $ $\text{hour}^{-1}$, $k_{2}=0.15 $ $\text{hour}^{-1}$, $k_{3}=0.45 $ $\text{hour}^{-1}$, and $k_{4}=0.06 $ $\text{hour}^{-1}$ for the two equations.

%The two angles, $\theta$ and $\varphi$ defined in FIG \ref{3}(b), involved in cap-like L$_\text{o}$ domains in micro phase were measured to be $60\pm20^\circ$ and $10\pm5^\circ$, respectively (FIG. \ref{3}). 

%\section{Discussions }

Here, phase separation dynamics of an asymmetric two-component vesicle were numerically calculated using purely dissipative dynamical model.
The vesicle is represented by an one-dimensional closed line in a two-dimensional plane, parameterized by $s$. The order parameter $\phi$ is defined as the local difference between the concentration of DPPC, $\phi_\text{DPPC}$, and that of DOPC, $\phi_\text{DOPC}$, per unit area, i.e., $\phi=\phi_\text{DPPC}-\phi_\text{DOPC}$. Additionally, for simplicity, we do not consider diffusion process of GM1 inserted in the outer layer. In terms of $\phi$, the two-component vesicle has the free energy functional $F=F_\text{1}+F_\text{2}$,
\begin{equation}
F_\text{1}=\int ds \frac{\kappa(\phi )}{2}\bigl( H-H_\text{sp}(\phi ) \bigr) ^2+pS,
\label{elastic} 
\end{equation}
 
\begin{equation}
F_\text{2}=\frac{k_\text{B}T}{a}\int ds \biggl[\frac{\xi_{0}^{2}}{2}\biggl(\frac{\partial \phi }{\partial s} \biggr) ^{2}+f(\phi )\biggr],
\label{phase separation} 
\end{equation}

%\begin{equation}
%F_\text{3}=pS.
%\label{osmotic pressure} 
%\end{equation}
$F_\text{1}$ is the sum of the bending elastic and osmotic energies where $\kappa (\phi )$ is the bending elastic modulus, $H/2$ the mean curvature and $H_\text{sp}(\phi )$ the spontaneous curvature, $p$ is the pressure difference across the vesicle and $S$ the enclosed area. Note that the Gaussian curvature is zero because we assume translation symmetry parallel to the z-axis.
We assume the following functional forms for $\kappa (\phi )$ and 
$H_\text{sp}(\phi )$: $\kappa (\phi )=\kappa_{0}+\kappa_{1}\phi $ and $H_\text{sp}(\phi )=H_\text{sp}^{(0)}+H_\text{sp}^{(1)}\phi$, where $\kappa_{0}$, $\kappa_{1}$, 
$H_\text{sp}^{(0)}$ and $H_\text{sp}^{(1)}$ are constants.  $F_{2}$ is the Ginzburg-Landau free energy where $k_\text{B}$ is the Boltzmann constant, $T$ the temperature, 
$a$ the unit cell length, $\xi_{0}$ the correlation length. The potential $f(\phi)$ is $a_{2}\phi^{2}/2+a_{4}\phi^{4}/4$, where $a_{2}$ and $a_{4}$ are the negative and positive coefficients, respectively.

\begin{figure}[htbp]
\centering
\includegraphics[width=80mm]{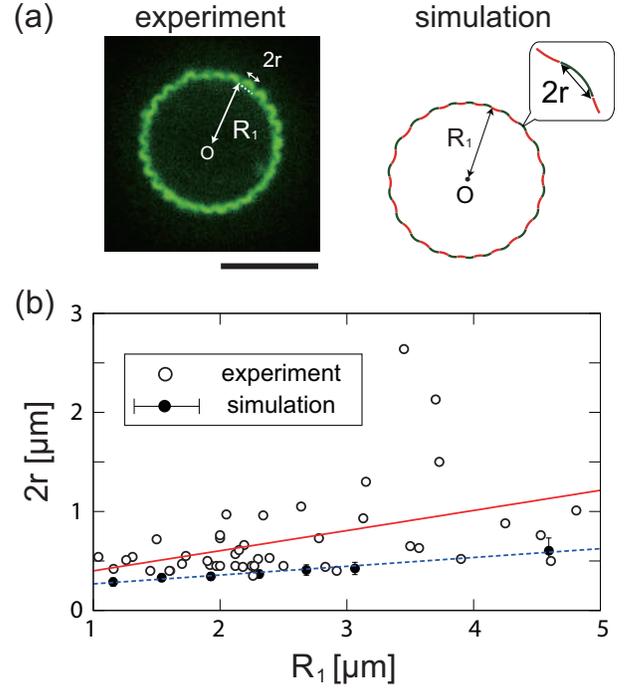}
\caption{\label{5}{ Size distribution of circular micro domains.} (a) The definitions of the domain size, $2r$, and the L$_\text{d}$ matrix radius, $R_\text{1}$, in the experiment and simulation. The scale bar is 10 $\mu$m. The domains where $\phi\geq 0$ (rich in DPPC) and $\phi<0$ (rich in DOPC) are shown in green and red respectively. (b) Domain size, $2r$, as a function of the L$_\text{d}$ matrix radius, $R_\text{1}$. Both experiments (open circle) and simulations (solid circles) are in excellent agreement with the linear least-squares fits, which are the red solid-line and blue dashed-line respectively. The slopes and intercepts in experiments and simulations are 0.20 $\pm$ 0.06 and 0.09 $\pm$ 0.18 $\mu$m, respectively. In the simulations, the vesicle size is varied from 1.16 $\mu$m to 4.56 $\mu$m, corresponding to the change of $N$ from 750 to 3000, for $\widetilde{\kappa}_{0}=3000$, $\widetilde{\kappa}_{1}=1/9$, $\widetilde{H}_\text{sp}^{(0)}=0.0765$, and $\widetilde{H}_\text{sp}^{(1)}=0.0075$. 
}
\end{figure}

We consider the time evolution of phase-separation transitions accompanied by dissipation of the free energy $F$. 
The position vector of a material point $s$ at a time $t$ is shown by $\mbox{\boldmath $r$}(s,t)=(x(s,t)$, $y(s,t))$. The simplest dissipative model reads
\begin{equation}
\frac{\partial \mbox{\boldmath $r$}(s,t)}{\partial t}=-\Gamma\frac{\delta }{\delta \mbox{\boldmath $r$}(s,t)}\biggl[F+\int \sqrt{g} ds \gamma(s,t) \biggr],
\label{TDGL} 
\end{equation}
where $\Gamma$ is a kinetic coefficient and $\sqrt{g}ds$ the line element. The incompressibility condition, $\sqrt{g}=\sqrt{(dx/ds)^{2}+(dy/ds)^{2}})=1$, for the local line element is guaranteed by a local Lagrange multiplier $\gamma(s,t)$~\cite{raft8}.
The following equations are derived from Eqs. (\ref{elastic})-(\ref{TDGL}).
See supplementary materials for the differential equations for $x$ and $y$.
The equation for $\gamma(s,t)$ is derived by substituting the differential equations of $x$ and $y$ in the relation $dg/dt=0$ as follows: 
$\ddot{\gamma}-H^{2}\gamma -D=0$ with $D=pH-\dot{x}\ddot{A_\text{y}}+\dot{y}\ddot{A_\text{x}}$ and $A_{\alpha}\equiv \dot{B}\dot{\alpha}+2B\ddot{\alpha}$  
where the dot on a variable $X$ denotes $\partial X/\partial s$ and $B$ is defined as $\kappa(\phi ) [H-H_\text{sp}(\phi )]$. 
\begin{figure}[ht!bp]
\centering
\includegraphics[width=80mm]{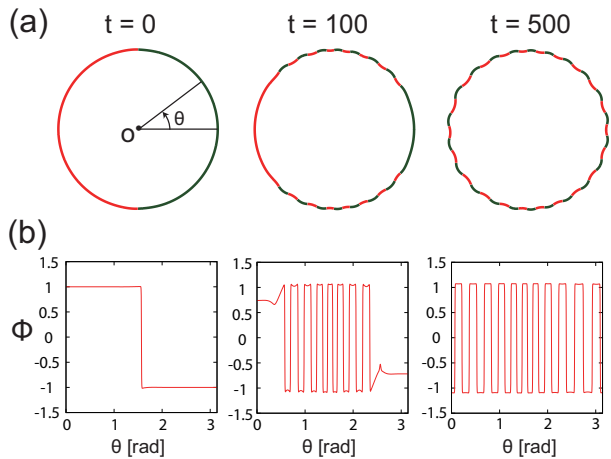}
\caption{\label{5}{ Phase separation dynamics after setting a positive spontaneous curvature.} (a) Cross sectional 1D vesicles at $t$=0, 100 and 500 in the phase-separation transitions. The domains where $\phi\geq 0$ (rich in DPPC) and $\phi<0$ (rich in DOPC) are shown in green and red respectively. (b) Periodic profile of $\phi$ as a function of the angle $\theta$. The simulation parameters are $N=2000$, $\widetilde{\kappa}_{0}=3000$, $\widetilde{\kappa}_{1}=1/9$, $\widetilde{H}_\text{sp}^{(0)}=0.0765$ and $\widetilde{H}_\text{sp}^{(1)}=0.0075$.
}
\end{figure}

Secondly, we consider phase separation in lipid bilayers. The variation of $\phi(s,t)$ per unit time is given by the equation of continuity, i.e., $\partial \phi(s,t)/\partial t=-(\partial/\partial s)J(s,t)$ where $J$ is the diffusion flux. The flux $J$ is proportional to the lateral gradient of the local chemical potential $\mu(s,t)$, namely $J=-L(\partial \mu(s,t)/\partial s)$, where $L$ is the transport coefficient. The chemical potential $\mu$ is given by $\delta F/\delta \phi(s,t)$ and thus is calculated as follows:
\begin{equation}
\mu(s,t)=\frac{k_\text{B}T}{a}\Bigl[-\xi_{0}^{2}\frac{\partial^{2}\phi}{\partial s^{2}}+\frac{\partial f}{\partial \phi}\Bigr]+\frac{1}{2}\frac{\partial \kappa }{\partial \phi } (H-H_\text{sp})-B\frac{\partial H_\text{sp}}{\partial \phi}. 
\label{chemical} 
\end{equation} 
In our simulation, space, time, and energy are non-dimensionalized using $\xi^{2}=\xi_{0}^{2}/|a_{2}|$, $a\xi^{2}/k_\text{B}TL|a_\text{2}|$ and $|a_\text{2}|\phi_\text{eq}^{2}k_\text{B}T\xi^{2}/a$, respectively, 
where $\phi_\text{eq}$ is defined as $\sqrt{|a_\text{2}|/a_\text{4}}$. 
This non-dimensionalization results in generating two scaled bending coefficients, 
$\widetilde{\kappa}_\text{0}=\kappa_{0}a/a_\text{2}\phi_\text{eq}^{2}k_\text{B}T\xi^{2}$ and 
$\widetilde{\kappa}_\text{1}=\kappa_\text{1}\phi_\text{eq}/\kappa_\text{0}$,
and two scaled spontaneous curvatures, $\widetilde{H}_\text{sp}^{(0)}=H_\text{sp}^{(0)}\xi$ and 
$\widetilde{H}_\text{sp}^{(1)}=H_\text{sp}^{(1)}\phi_\text{eq}\xi$.
We have attempted to use experimental results for simulation parameters.
The circular vesicle radius at the initial state is estimated by $N\xi/2\pi$, where $N$ is a material point number.   
Thus, for example, by substituting $N=2000$ and a reasonable assumption $\xi=10$ $\text{nm}$ in this relation, 
the vesicle radius is obtained to be 3.2 $\mu\text{m}$, which is close to the average vesicle radius observed in the experiments (FIG. S1).
The bending coefficient $\widetilde{\kappa}_\text{0}$ is difficult to determine from experimental values and it is the only undetermined parameter. 
On the other hand, the bending coefficient $\widetilde{\kappa}_\text{1}$ is estimated to be 1/9 by using the ratio of the bending coefficient in DPPC rich phase to that in DOPC rich phase, 1.25~\cite{webb}. 
The spontaneous curvature of the bilayer in each phase is obtained by the relationship $H_\text{sp}=(H_\text{sp}^\text{out}-H_\text{sp}^\text{in})/2$, where 
$H_\text{sp}^\text{in}$ and $H_\text{sp}^\text{out}$ are the spontaneous curvatures of the inner and outer monolayers~\cite{spon1, spon2}. 
We assume that the spontaneous curvature of a monolayer is the linear function of the mole fractions of components~\cite{mole fraction1, mole fraction2}.  
Using the reported spontaneous curvatures of DPPC, DOPC and GM1 (0.003 $\text{\AA}^{-1}$~\cite{DPPC}, $-0.006$ $\text{\AA}^{-1}$~\cite{mole fraction1} and 0.017 $\text{\AA}^{-1}$~\cite{GM1}, respectively) and the 
measured mole fractions of GM1 in two phases ($\sim$ 6\% in L$_\text{o}$ phase and 
$\sim$ 3\% in L$_\text{d}$ phase), the spontaneous curvatures of the bilayers in two phases are obtained.
Using the above values, $\widetilde{H}_\text{sp}^{(0)}$ and $\widetilde{H}_\text{sp}^{(1)}$ are estimated to be 0.0765 and 
0.0075.
In our model, we have ignored the flip-flop phenomena of GM1 having a bulky and hydrophilic head group as even smaller lipids without such a 
head group show very slow reaction rates (several hours to a day)~\cite{flip}.

%the difference of the amount of amphiphiles diffusing into and out of the line element. Thus, time variation for $\phi(s,t)$ satisfies 

%The simulation parameters are the material point number $N$, non-dimensionalized bending coefficients 
%, and non-dimensionalized spontaneous curvatures $\widetilde{H_\text{sp}^{(0)}}=H_\text{sp}^{(0)}\xi$ and $\widetilde{H_\text{sp}^{(1)}}=H_\text{sp}^{(1)}\phi_\text{eq}\xi$,
%where $\xi$ and $\phi_\text{eq}$ are defined as $\xi_{0}/a_\text{2}$ and $\sqrt{|a_\text{2}|/a_\text{4}}$, respectively.  because we have set the length between two neighboring material points as $\xi$. 

In our simulation, we set fully phase-separated circular vesicle with zero spontaneous curvature, i.e., $\widetilde{H}_\text{sp}^{(0)}=0$ and $\widetilde{H}_\text{sp}^{(1)}=0$. 
The vesicle was sufficiently relaxed to an equilibrium state. 
At $t=0$ we assume that GM1 was instantly inserted into the outer layer, which results in a change of the spontaneous curvature in each domain.
We set the non-zero spontaneous curvatures, $\widetilde{H}_\text{sp}^{(0)}=0.0765$ and $\widetilde{H}_\text{sp}^{(1)}=0.0075$, 
and observed the following dissipative process from $t=0$ to $t=500$.   
Note that the instant insertion of GM1 is reasonable as the time scale of insertion is negligible compared to that of the subsequent phase separation processes. Based on the framework, we performed numerical simulations using the following parameters: $N=2000$, $\widetilde{\kappa}_{0}=3000$, $\widetilde{\kappa}_{1}=1/9$, $\widetilde{H}_\text{sp}^{(0)}=0.0765$ and $\widetilde{H}_\text{sp}^{(1)}=0.0075$; so that we observed the gradual phase separation transitions with shape deformations (FIG. 4, Movie S4).   
In FIG. 4(a), the domains where $\phi\geq 0$ (rich in DPPC) and $\phi<0$ (rich in DOPC) are shown in green and red respectively. 
The monodisperse L$_\text{o}$ domains are distributed at approximately regular intervals along the line and slightly bud toward the outside of the vesicle, 
which are in excellent agreement with experimental results (FIG. 4).
Qualitatively speaking, the transition results in the increase of the boundary energy $F_{2}$, but also the decrease of the bending elastic energy, $F_{1}$, as the budding of L$_\text{o}$ phase causes the membrane curvature to become a closer value to the spontaneous curvature, therefore total free energy decreased.
Also, note that the bending energy is decreased more by budding of the stiffer L$_\text{o}$ domains both in experiment and numerical simulation.
Moreover, we compared $R_\text{1}$ and $2r$ obtained from our experimental results with those from our numerical calculations. 
The value $R_\text{1}$ has been defined as the average length between the center of the circular vesicle and each material point which satisfies $\phi<0$ (FIG. 3(a)).
The value $2r$ has been defined as the average length between two neighboring material points at which the sign of $\phi$ is changed as a function of $\theta$ (FIG. 3(a)).  
The analyzed values and those standard deviations are plotted in FIG.3(b) with the best fitting (the blue dashed-line), 
where the material point number alone has been varied from $N=750$ to $N=3000$ 
for $\widetilde{\kappa}_{0}=3000$, $\widetilde{\kappa}_{1}=1/9$, $\widetilde{H}_\text{sp}^{(0)}=0.0765$, and $\widetilde{H}_\text{sp}^{(1)}=0.0075$.    
Quantitatively, the numerical results are mostly consistent with experimental results.
Furthermore, we investigated a domain number $N_\text{cap}$, $2r/R_{1}$ and total free energy as a function of the mole fraction of GM1 in L$_\text{o}$ phase. The ratio of the mole fraction in L$_\text{o}$ phase to that in L$_\text{d}$ for $N=1000$, $\widetilde{\kappa}_{0}=3000$ and $\widetilde{\kappa}_{1}=1/9$, which is kept constant, equals to two.
We have found that there are discontinuous jumps of $N_\text{cap}$ and $2r/R_{1}$ as a function of the mole fraction.
Although $N_\text{cap}=1$ below the transition fraction, indicating a vesicle of macro-phase, $N_\text{cap}$ increases and $2r/R_{1}$ decreases as the mole fraction of GM1 increases above it (FIG. S6).
However, the detail of the discontinuous transition remains poorly understood.

%after the insertion of GM1 is about one order of magnitude larger than the membrane curvature before it 
%The profiles of $\phi$ in FIG. 5(b) show that the two phases are a little more strongly phase separated after transitions.
%We now describe the procedure of the numerical calculation we have performed. 

%After changing the positive curvature, we have observed the gradual phase separation transitions with shape deformations for $N=2000$, $\kappa_{0}=3000$, $\kappa_{1}=1/9$, 
%$H_\text{sp}^{0}=0.0765$ and $H_\text{sp}^{1}=0.0075$ (FIG. 5).
%The line color difference in FIG.5 (a) was determined by the condition $\phi\geq 0$ (green, DPPC rich region) or $\phi<0$ (red, DOPC rich region).

To summarize, we have experimentally found that fully phase-separated phase with asymmetric lipid composition undergoes a transition to the micro phase 
via the lorate-phase as a metastable state, resulting in the emergence of monodisperse submicron scale micro domains. 
Numerical analysis based on the bending elastic model and TDGL equation shed some light on the mechanism of the transitions, implying the significance of the spontaneous curvature due to the asymmetric composition in the thermodynamic stabilization of micro-phase 
separation in lipid bilayers.  
Our findings could give some important clues as to the physical principle for new material designs and behind the heterogeneities existing in cell membranes. 

We thank Professor K. Yoshikawa for useful suggestions. This work was supported by Grant-in-Aid for JSPS Fellows Grant (No. 25-1270) and by KAKENHI (Nos. 26707020, 25103012 and 26115709).

\end{document}